\documentclass[11pt]{aa}  

\usepackage{graphicx}
\usepackage{txfonts}
\usepackage{soul}

\usepackage[colorlinks,citecolor=blue,urlcolor=blue]{hyperref}

\begin{document} 

    \title{Unexpected frequency of horizontal oscillations of magnetic structures in the solar photosphere}

    \author{M. Berretti \inst{1,2}
            \and
            M. Stangalini \inst{3}
            \and
            G. Verth \inst{4}
            \and
            S. Jafarzadeh \inst{5,6}
            \and
            D.~B. Jess \inst{7,8}
            \and
            F. Berrilli \inst{2}
            \and
            S.~D.~T. Grant \inst{7}
            \and
            T. Duckenfield \inst{7}
            \and
            V. Fedun \inst{9}
            }

    \authorrunning{Berretti et al.}

   \institute{University of Trento,
              Via Calepina 14, 38122 Trento, Italy,
              \email{michele.berretti@unitn.it}
         \and
             University of Rome Tor Vergata, Department of Physics, Via della Ricerca Scientifica 3, 00133 Rome, Italy
         \and 
             ASI Italian Space Agency, Via del Politecnico snc, 00133 Rome, Italy
          \and
            Plasma Dynamics Group, School of Mathematics and Statistics, The University of Sheffield, Hicks Building, Hounsfield Road, Sheffield, S3 7RH, UK
          \and
             Max Planck Institute for Solar System Research, Justus-von-Liebig-Weg 3, 37077 G\"{o}ttingen, Germany
          \and
          Niels Bohr International Academy, Niels Bohr Institute, Blegdamsvej 17, DK-2100 Copenhagen, Denmark
          \and 
             Astrophysics Research Centre, School of Mathematics and Physics, Queen’s University Belfast, Belfast, BT7 1NN, Northern Ireland, UK
            \and
            Department of Physics and Astronomy, California State University Northridge, Northridge, CA 91330, USA
             \and
            Plasma Dynamics Group, Department of Automatic Control and Systems Engineering, The University of Sheffield, Mappin Street, Sheffield, S1 3JD, UK
}

   \date{}

  \abstract
  {It is well known that the dominant frequency of oscillations in the solar photosphere is $\approx$3~mHz, which is the result of global resonant modes pertaining to the whole stellar structure. However, analyses of the horizontal motions of nearly $1$~million photospheric magnetic elements spanning the entirety of solar cycle 24 have revealed an unexpected dominant frequency, $\approx$5~mHz, a frequency typically synonymous with the chromosphere. Given the distinctly different physical properties of the magnetic elements examined in our statistical sample, when compared to largely quiescent solar plasma where $\approx$3~mHz frequencies are omnipresent, we argue that the dominant $\approx$5~mHz frequency is not caused by the buffeting of magnetic elements, but instead is due to the nature of the underlying oscillatory driver itself. This novel result was obtained by exploiting the unmatched spatial and temporal coverage of magnetograms acquired by the Helioseismic and Magnetic Imager (HMI) on board NASA's Solar Dynamics Observatory (SDO). Our findings provide a timely avenue for future exploration of the magnetic connectivity between sub-photospheric, photospheric, and chromospheric layers of the Sun's dynamic atmosphere. }


   \maketitle

    \section{Introduction}

    It is widely acknowledged that magnetic concentrations are present across the entire surface of the Sun and can reach dimensions down to the spatial limit of current high-resolution observatories \citep{lagg_fully_2010, rutten_solar_2020}. These small-scale magnetic structures are anchored in the photosphere and permeate upwards through the different layers of the solar atmosphere \citep{stenflo_small-scale_1989}. Consequently, they are considered key elements in the energetic balance of the Sun's atmosphere. 
    
    Given the photospheric plasma-$\beta$, small-scale magnetic elements are frozen into the turbulent photospheric plasma and are dominated by the gas pressure, and are hence subjected to a diffusion process bound to the plasma flows. Numerous works have studied the horizontal motions of bright points in the photosphere, describing their diffusive dynamics \citep{1999ApJ...511..932H, 2003ApJ...587..458N, 2011ApJ...743..133A, 2012ApJ...752...48C}.  \citet{giannattasio_diffusion_2013} show the displacement spectrum of more than 20{\,}000 magnetic elements in the photosphere and report a super-diffusive process with a coefficient that depends on the scale of the elements considered. Moreover, \citet{jafarzadeh_migration_2014} studied the migration of small-scale magnetic features in the photosphere (diameters of around $0{\,}.{\!\!}''2$) and interpreted the super-diffusive process as the resulting superposition of turbulent granular evolution and more steady flows on meso-granular and super-granular scales. Therefore, these magnetic structures can be considered an excellent proxy of the driving mechanisms within the photosphere, whose kinematics depend on the plasma environment they are embedded in \citep{2017ApJS..229....8J}.
    
   \begin{figure*}[h!]
       \centering
       \includegraphics[]{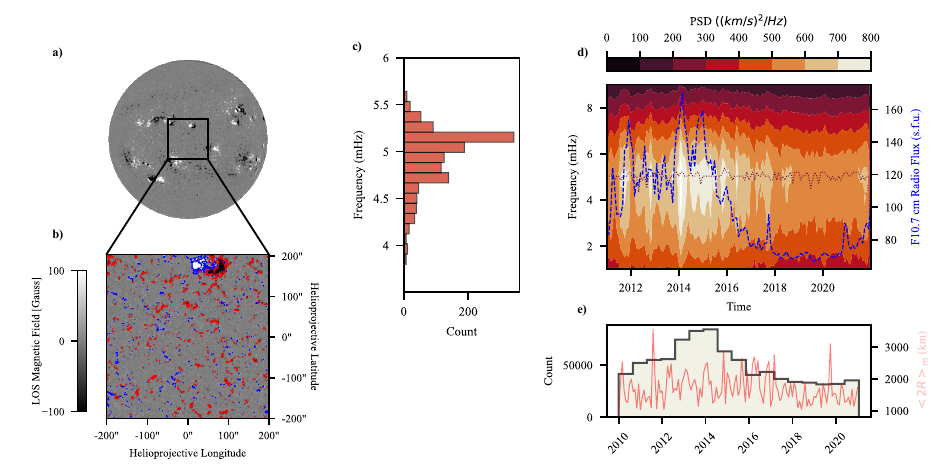}
       \caption{\textbf{Analysis of the horizontal velocity oscillations of the magnetic structures considered. (a)} Sample full-disk magnetogram obtained by SDO/HMI. \textbf{(b)} Field of view considered in this analysis. Both magnetograms are saturated between -100~G and 100~G. The boundaries of elements detected by the feature-tracking algorithm for positive and negative polarities are contoured using blue and red lines, respectively. \textbf{(c)} Statistical distribution of the dominant frequency within each observational window. The displayed frequency range is constrained between $3.5$ and $6.5$~mHz. No significant values are observed outside this range. \textbf{(d)} Results of the time-frequency analysis of the elements' horizontal velocities ($v_h$). The dotted brown line highlights the dominant frequency during each month. The blue line represents the observed radio flux, a common physical proxy used to track the solar activity cycle. \textbf{(e)} Statistical distribution of the number of magnetic structures detected and tracked during the solar activity cycle. The red line represents the average equivalent diameter of the observed magnetic concentrations during each month ($<2R>_m$).}
       \label{fig:1}
   \end{figure*}
   
    These magnetic fields are brought to the surface by convective upflows. Subsequently, they are passively advected towards the boundaries of granular cells \citep{stenflo_magnetic-field_1973, jess_rosa_2010, riethmuller_comparison_2014, borrero_solar_2017}. Within the intergranular lanes, small-scale magnetic structures are continuously swayed by the movements of neighbouring granules \citep{steiner_dynamical_1998, hasan_excitation_2000}, leading to horizontal displacements regarded as the combination of coherent oscillations and random walk \citep{roberts_2019}. Furthermore, the magnetic structures are subjected to repeated `kicks' from changes in the nearby granulation pattern (i.e. either emerging or merging or exploding granules). This can lead to a number of wave modes excited within the magnetic structure \citep{edwin_roberts_1982, edwin_wave_1983}. These oscillations can propagate through the outer layers of the solar atmosphere, acting as probes that reveal local plasma conditions and helping to answer long-debated questions in solar physics regarding, for example, coronal heating \citep{Goossens_2013, vanDoorsselaere2014, van_doorsselaere_coronal_2020} and solar wind acceleration \citep{de-pontieu_chromospheric_2007}.
    
    Considering the nature of the photospheric driver, we expect these periodic oscillations to show frequencies compatible with those usually observed in the photosphere \citep[see e.g. the recent reviews by][]{2015SSRv..190..103J, 2023LRSP...20....1J}. The photosphere is known to be dominated by either acoustic oscillations at $\approx$3~mHz, which are the result of global $p$-modes (i.e. the global resonant mode of the whole stellar structure) or the lifetime of granulation at $1.6-2.0$~mHz \citep{delmoro_granulation-2004, centeno_emergence_2007}. In this regard, \citet{stangalini_observational_2014} studied the horizontal oscillations of 22 small-scale magnetic elements in the photosphere using empirical mode decomposition techniques to account for the non-stationary traits of the time series. In their analysis, they show observational evidence of kink waves in the magnetic elements, which were excited by the continuous buffeting of the granules with a dominant period compatible with the lifetime of granulation. 
    
    The Helioseismic and Magnetic Imager \citep[HMI;][]{schou_design_2012}, the magnetograph on board NASA's Solar Dynamics Observatory \citep[SDO;][]{pesnell_solar_2012}, has been providing continuous observations of magnetic concentrations in the photosphere for more than 10~years. In this work, we exploit the availability of stable and seeing-free HMI data to provide the most comprehensive statistical analysis to date of the coherent (i.e. periodic) oscillations in the horizontal velocity of small-scale magnetic elements in the lower solar atmosphere. This resulted in more than 30~million magnetic elements detected and tracked during the whole operational lifetime of the instrument.

    \section{Dataset and analysis}

    We analysed magnetograms acquired with SDO/HMI in the Fe~{\sc i}~617.3 nm absorption line, with a cadence of 45 seconds, in observational windows of 40 minutes each, every three days from January 1, 2011, to November 29, 2021. This dataset consists of around 11 years of observations of the line-of-sight component of the photospheric magnetic fields spanning the full solar cycle 24 over a $400\times400$~arcsec$^{2}$ patch located at the centre of the solar disk (see Fig. \ref{fig:1}). The sequences of magnetograms taken during each observational window are co-aligned in order to account for the solar rotation (i.e. the field of view is fixed, and the different frames are co-registered to each other). The Southwest Automatic Magnetic Identification Suite \citep[SWAMIS;][]{deforest_solar_2007} feature-tracking algorithm was then executed on the aligned magnetograms. 
    
    SWAMIS works by comparing pixels in a given image with two thresholds, a higher one to detect the coarse positions and borders of magnetic concentrations, and a lower one for finer adjustments of the detected borders. Specifically, pixels in an image are compared to a high threshold and marked. Subsequently, all pixels neighbouring the marked ones are compared to the lower threshold. The neighbourhood is defined as a $3\times3\times3$ cube in both space and time. Thus, the detected clusters are dilated until they no longer exceed the lower threshold. This approach ensures hysteresis in both space and time, further boosting confidence in the detected elements.
    The higher threshold was set to 6$\sigma$ above the noise, and the lower one to 2$\sigma$. The value of $\sigma$ was inferred by averaging the root-mean-square of a quiet-Sun patch in the magnetograms over each temporal window and was found to equal $\approx$7~G. This overestimates the actual polarimetric noise that would be obtained through continuum measurements, possibly leading to a number of undetected magnetic elements that can be considered unimportant thanks to the already exceptional size of the dataset. Moreover, despite the lower threshold corresponding to the 95\% confidence level, the constraints imposed on the detection algorithm regarding the minimum size and lifetime of the elements (e.g. no fewer than four pixels based on the full width at half maximum of the instrument and no fewer than four consecutive time steps) guarantee an actual confidence level much higher than 95\%. In this work, we are interested in the instantaneous horizontal velocities of the detected elements, which were obtained through the first-order derivative of the positions of the barycentres of the detected structures over time. Barycentres are found by averaging the coordinate of each pixel belonging to a feature in a magnetogram, weighted by their intensity; this results in sub-pixel accuracy.
    
    In panel A of Fig.\ref{fig:1} we show a sample full-disk magnetogram as captured by HMI. In panel B we show a sample of the field of view considered in this work and the results of the tracking algorithm.

    \section{Results}
    To achieve sufficient frequency resolution for our analysis, we limited our dataset to only elements that survived for more than 30 consecutive time steps (i.e. more than 22~minutes), resulting in just below 1~million magnetic elements considered for the spectral analysis (precisely 851{\,}912, i.e. an order of magnitude less than the total number of detected structures). Furthermore, to obtain reliable power spectral densities from fast Fourier transform analysis, the signals required a number of pre-processing steps \citep[see][]{2023LRSP...20....1J}. First, we removed any linear background trends that could be present in the time series by subtracting the result of a linear least-squares fit (detrending). To avoid issues caused by the truncation of the series, we applied a Tukey window function with $\alpha=0.5$ (apodising) and zero-padded the time series to 128 to set their lengths to a power of two (to increase computational performance) and increase the display resolution. Finally, we computed the associated fast Fourier transforms.

    \begin{figure}[h!]
        \centering
        \includegraphics[width=\hsize]{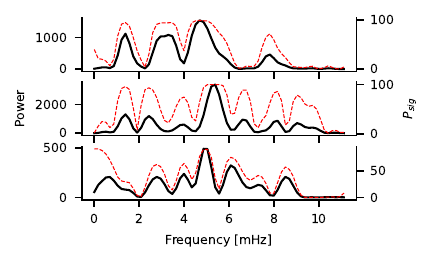}
        \caption{Three periodograms depicting the normalised power spectral density of the $x$ component of the horizontal velocity ($v_{h,x}$) of the magnetic tubes (solid black line) randomly picked  from the dataset. The dashed red lines are the percentage probabilities computed over 1500 randomised variations of the input $v_{h,x}$.}
        \label{Fig2}
    \end{figure}
    
    To determine the confidence level shown in Fig. \ref{Fig2}, we followed the procedure highlighted in \citet{2023LRSP...20....1J}. We obtained the probability, $p_{sig}$, that the observed oscillatory phenomenon is significant by comparing the periodogram of the input time series with 1500 periodograms of randomised series based on the original signal. $p_{sig} = 99\%$ corresponds to a confidence level of 95\% \citep{1985AJ.....90.2317L, 2001A&A...371.1137B}. While this process can be helpful to showcase the reliability of the power spectra in a few samples, it is practically impossible to apply it to all the elements in the dataset given its size. Moreover, this method is used here for illustrative purposes. Given the substantial size of our dataset, a more appropriate approach would be to compute the confidence level using the standard deviation and the sample size.

    In panel D of Fig. \ref{fig:1} we show the time-frequency diagram of the horizontal velocity oscillation. Inspired by the B-$\omega$ diagram introduced in \citet{stangalini_novel_2021}, it shows the average spectral density of the observed features over each month. It has been adapted to consider multiple small-scale magnetic elements rather than single pixels of a magnetic pore or sunspot. Panel D of Fig. \ref{fig:1} reveals a frequency band with dominant power centred between $2$ and $6$~mHz. Furthermore, it shows a clear correlation between the width and power of the frequency band and the solar activity cycle. In panel E of Fig. \ref{fig:1} we show the number of magnetic structures detected over time. Interestingly, during periods of minimum solar activity, the number of small-scale magnetic elements detected remains constant. Conversely, the monthly averaged equivalent diameter (red line in panel E of Fig. \ref{fig:1}) shows a slight trend towards higher concentrations during the solar maximum, likely linked to the global enhancement of all magnetic activity.

    We then consolidated the normalised power spectral densities into a single, averaged spectral power distribution that incorporated every observational window. The aim of the consolidation process was to simplify the analysis and allow for a statistical study of significant common trends found in such an exceptionally large dataset.
    Moreover, normalising the spectra prior to averaging them together is necessary as we are mainly concerned with the position of the frequency peak and its relative amplitude, rather than its absolute power. Otherwise, elements with much stronger spectral amplitudes would bias the actual position of the peak in the ensemble.
    
    In panel C of Fig. \ref{fig:1}, we show the histogram of the dominant frequency of the horizontal velocities ($v_h$) in each observational window (i.e. the frequency peak with the highest power in the periodograms). There is a clear peak centred at $\approx$5~mHz (3~minutes). This is an unexpected result as 3-minute periodicities are not commonly observed elsewhere at photospheric heights and are far more frequently observed in the chromosphere. Except for a few outliers, the majority of elements have a lifetime exceeding 25 minutes. This allows us to confidently assert the robustness of the identified 3-minute periodicity since, on average, each time series spans more than eight cycles, thus providing a solid basis for the observed oscillations. 

    To verify the consistency of the methods used in this work and the genuineness of our results, we carried out an alternative analysis using the wavelet transform instead of the Fourier transform (see Appendix \ref{apb}).  The same results are obtained.

    \section{Discussions}

    The broad band shown in the time-frequency diagram in the top panel of Fig. \ref{fig:1} highlights the presence of a unique series of peaks in the power spectra of the horizontal velocity perturbations of each element. Given the high plasma-$\beta$ of the surrounding photosphere, the magnetic elements can therefore be considered passive tracers of the plasma flows. For this reason, we argue that the main peak at $5$ mHz as well as these additional frequencies are inherent to the photospheric driver. This is consistent with prior findings by \citet{1998ApJ...509..435V}, \citet{2013ApJ...768...17M}, \citet{stangalini_spectrum_2013}, \citet{jafarzadeh_high-frequency_2017}, \citet{jess_inside_2017}, \citet{keys_pores_2018}, \citet{2020NatAs...4..220J}, \citet{stangalini_novel_2021}, and \citet{grant_pores_2022}, to name but a few examples of studies of photospheric dynamics driving wave and oscillatory motions in the lower solar atmosphere. 
    
    Conversely, the distribution of the dominant frequency in each observational window shows a narrow peak at 5~mHz. Understanding the nature of the observed $5$~mHz frequency is a challenging task. This frequency is not in agreement with two typical timescales in the photosphere, namely the global $p$-modes (i.e. $3$ mHz) and the typical lifetime of granules (i.e. $1.6-2$~mHz). However, since the $5$ mHz oscillation is found in a sample of different magnetic elements  (in terms of size and flux), we argue that this can be the result of a global process (i.e. one not linked to the local environmental conditions of the plasma). It is worth noting that the amplitude of these oscillations is also modulated with the solar activity cycle, which further supports our argument. We also note that the absence of a frequency modulation with the solar cycle is not in agreement with a possible variation of the granular scales \citep{2007A&A...475..717M}, and therefore we speculate that this could be linked to a sub-surface driving process whose amplitude changes with the solar dynamo. Interestingly, given the amplitude of the horizontal velocity oscillations as inferred by the tracking, the observed 5 mHz frequency is compatible with the crossing time of intergranular lanes. However, one possible limitation of this interpretation is revealed by the dimensions of the flux tubes: Most of them have dimensions of the order of a thousand kilometres, as shown in Appendix \ref{apa}, which is compatible with the dimensions of granules. It remains unclear how to reconcile the compatibility between the crossing time of intergranular lanes and the dimensions of these elements. Finally, it is worth noting that the same analysis was carried out by filtering out frequencies lower than 2~mHz, which are most likely linked to slow evolution rather than waves, and the same results are obtained.

    Despite the novelty of our results, it is important to note that they are consistent with previous literature. It has been demonstrated that a power law behaviour with a frequency peak is possible \citep[see][]{2009ApJ...697.1384T}. However, many earlier studies did not observe the 5 mHz peak. We believe that this discrepancy could be due to the limited number of elements considered and their shorter average lifetimes when compared to those in our dataset, or because previous similar analyses did not specifically look for this peak (e.g. potentially excluding it by selecting only subsonic contributions).

    \section{Conclusions}

    In this work we have studied the horizontal perturbations of nearly 1 million small-scale magnetic elements as observed in magnetogram sequences taken by HMI/SDO in the Fe~{\sc i} 617.3~nm absorption line. The elements were detected and tracked using the SWAMIS suite in a $400\times400$~arcsec$^{2}$ region located at disk centre in the photosphere and spanning a full solar cycle. We show a high power frequency band centred at 5~mHz that is shared across the vast majority of the elements in the dataset and is stable throughout the whole solar cycle. This is an unexpected result, as frequencies usually observed in the photosphere are more commonly associated with either $p$-modes or the lifespan of granulation. However, this novel photospheric frequency has no immediate correspondence with typical timescales in the photosphere.

    On the other hand, a 5mHz dominant frequency is more commonly found in the chromosphere. Therefore, our next goal would be to investigate the magnetic link between the photosphere and the chromosphere, as well as determine the origin of such a frequency at photospheric heights.

    \begin{acknowledgements}
    We wish to acknowledge scientific discussions with the Waves in the Lower Solar Atmosphere (WaLSA; \href{https://WaLSA.team}{www.WaLSA.team}) team, which has been supported by the Research Council of Norway (project no. 262622), The Royal Society (award no. Hooke18b/SCTM), and the International Space Science Institute (ISSI Team 502). MB acknowledges that this publication (communication/thesis/article, etc.) was produced while attending the PhD program in  PhD in Space Science and Technology at the University of Trento, Cycle XXXIX, with the support of a scholarship financed by the Ministerial Decree no. 118 of 2nd March 2023, based on the NRRP - funded by the European Union - NextGenerationEU - Mission 4 "Education and Research", Component 1 "Enhancement of the offer of educational services: from nurseries to universities” - Investment 4.1 “Extension of the number of research doctorates and innovative doctorates for public administration and cultural heritage” - CUP E66E23000110001. MB also acknowledges the valuable discussions held with Simone Mestici.
    SJ acknowledges support from the Rosseland Centre for Solar Physics (RoCS), University of Oslo, Norway.
    DBJ and TD acknowledge support from the Leverhulme Trust via the Research Project Grant RPG-2019-371. DBJ and SDTG wish to thank the UK Science and Technology Facilities Council (STFC) for the consolidated grants ST/T00021X/1 and ST/X000923/1. DBJ and SDTG also acknowledge funding from the UK Space Agency via the National Space Technology Programme (grant SSc-009). VF and GV are grateful to the  Science and Technology Facilities Council (STFC) grants ST/V000977/1 and ST/Y001532/1. They also thank the Institute for Space-Earth Environmental Research (ISEE, International
Joint Research Program, Nagoya University, Japan), the Royal Society, International Exchanges Scheme, collaboration with Greece (IES/R1/221095), India (IES/R1/211123) and Australia (IES/R3/213012) for the support provided.
    \end{acknowledgements}

\bibliographystyle{aa}
\bibliography{bibcut}

\begin{thebibliography}{45}
\expandafter\ifx\csname natexlab\endcsname\relax\def\natexlab#1{#1}\fi

\bibitem[{{Abramenko} {et~al.}(2011){Abramenko}, {Carbone}, {Yurchyshyn}, {Goode}, {Stein}, {Lepreti}, {Capparelli}, \& {Vecchio}}]{2011ApJ...743..133A}
{Abramenko}, V.~I., {Carbone}, V., {Yurchyshyn}, V., {et~al.} 2011, \apj, 743, 133

\bibitem[{{Banerjee} {et~al.}(2001){Banerjee}, {O'Shea}, {Doyle}, \& {Goossens}}]{2001A&A...371.1137B}
{Banerjee}, D., {O'Shea}, E., {Doyle}, J.~G., \& {Goossens}, M. 2001, \aap, 371, 1137

\bibitem[{Borrero {et~al.}(2017)Borrero, Jafarzadeh, Schüssler, \& Solanki}]{borrero_solar_2017}
Borrero, J.~M., Jafarzadeh, S., Schüssler, M., \& Solanki, S.~K. 2017, ßr, 210, 275

\bibitem[{Centeno {et~al.}(2007)Centeno, Socas-Navarro, Lites, Kubo, Frank, Shine, Tarbell, Title, Ichimoto, Tsuneta, Katsukawa, Suematsu, Shimizu, \& Nagata}]{centeno_emergence_2007}
Centeno, R., Socas-Navarro, H., Lites, B., {et~al.} 2007, \apjl, 666, L137

\bibitem[{{Chitta} {et~al.}(2012){Chitta}, {van Ballegooijen}, {Rouppe van der Voort}, {DeLuca}, \& {Kariyappa}}]{2012ApJ...752...48C}
{Chitta}, L.~P., {van Ballegooijen}, A.~A., {Rouppe van der Voort}, L., {DeLuca}, E.~E., \& {Kariyappa}, R. 2012, \apj, 752, 48

\bibitem[{{De Pontieu} {et~al.}(2007){De Pontieu}, {McIntosh}, {Carlsson}, {Hansteen}, {Tarbell}, {Schrijver}, {Title}, {Shine}, {Tsuneta}, {Katsukawa}, {Ichimoto}, {Suematsu}, {Shimizu}, \& {Nagata}}]{de-pontieu_chromospheric_2007}
{De Pontieu}, B., {McIntosh}, S.~W., {Carlsson}, M., {et~al.} 2007, Science, 318, 1574

\bibitem[{DeForest {et~al.}(2007)DeForest, Hagenaar, Lamb, Parnell, \& Welsch}]{deforest_solar_2007}
DeForest, C.~E., Hagenaar, H.~J., Lamb, D.~A., Parnell, C.~E., \& Welsch, B.~T. 2007, The Astrophysical Journal, 666, 576

\bibitem[{{Del Moro}(2004)}]{delmoro_granulation-2004}
{Del Moro}, D. 2004, \aap, 428, 1007

\bibitem[{{Edwin} \& {Roberts}(1982)}]{edwin_roberts_1982}
{Edwin}, P.~M. \& {Roberts}, B. 1982, \solphys, 76, 239

\bibitem[{Edwin \& Roberts(1983)}]{edwin_wave_1983}
Edwin, P.~M. \& Roberts, B. 1983, \solphys, 88, 179

\bibitem[{Giannattasio {et~al.}(2013)Giannattasio, Moro, Berrilli, Rubio, Gos˘ić, \& Suárez}]{giannattasio_diffusion_2013}
Giannattasio, F., Moro, D.~D., Berrilli, F., {et~al.} 2013, The Astrophysical Journal Letters, 770, L36, publisher: The American Astronomical Society

\bibitem[{Goossens {et~al.}(2013)Goossens, Doorsselaere, Soler, \& Verth}]{Goossens_2013}
Goossens, M., Doorsselaere, T.~V., Soler, R., \& Verth, G. 2013, The Astrophysical Journal, 768, 191

\bibitem[{{Grant} {et~al.}(2022){Grant}, {Jess}, {Stangalini}, {Jafarzadeh}, {Fedun}, {Verth}, {Keys}, {Rajaguru}, {Uitenbroek}, {MacBride}, {Bate}, \& {Gilchrist-Millar}}]{grant_pores_2022}
{Grant}, S.~D.~T., {Jess}, D.~B., {Stangalini}, M., {et~al.} 2022, \apj, 938, 143

\bibitem[{{Hagenaar} {et~al.}(1999){Hagenaar}, {Schrijver}, {Title}, \& {Shine}}]{1999ApJ...511..932H}
{Hagenaar}, H.~J., {Schrijver}, C.~J., {Title}, A.~M., \& {Shine}, R.~A. 1999, \apj, 511, 932

\bibitem[{Hasan {et~al.}(2000)Hasan, Kalkofen, \& van Ballegooijen}]{hasan_excitation_2000}
Hasan, S.~S., Kalkofen, W., \& van Ballegooijen, A.~A. 2000, \apjl, 535, L67

\bibitem[{Jafarzadeh {et~al.}(2014)Jafarzadeh, Cameron, Solanki, Pietarila, Feller, Lagg, \& Gandorfer}]{jafarzadeh_migration_2014}
Jafarzadeh, S., Cameron, R.~H., Solanki, S.~K., {et~al.} 2014, \aap, 563, A101

\bibitem[{{Jafarzadeh} {et~al.}(2017{\natexlab{a}}){Jafarzadeh}, {Solanki}, {Cameron}, {Barthol}, {Blanco Rodr{\'\i}guez}, {del Toro Iniesta}, {Gandorfer}, {Gizon}, {Hirzberger}, {Kn{\"o}lker}, {Mart{\'\i}nez Pillet}, {Orozco Su{\'a}rez}, {Riethm{\"u}ller}, {Schmidt}, \& {van Noort}}]{2017ApJS..229....8J}
{Jafarzadeh}, S., {Solanki}, S.~K., {Cameron}, R.~H., {et~al.} 2017{\natexlab{a}}, \apjs, 229, 8

\bibitem[{{Jafarzadeh} {et~al.}(2017{\natexlab{b}}){Jafarzadeh}, {Solanki}, {Stangalini}, {Steiner}, {Cameron}, \& {Danilovic}}]{jafarzadeh_high-frequency_2017}
{Jafarzadeh}, S., {Solanki}, S.~K., {Stangalini}, M., {et~al.} 2017{\natexlab{b}}, \apjs, 229, 10

\bibitem[{{Jess} {et~al.}(2023){Jess}, {Jafarzadeh}, {Keys}, {Stangalini}, {Verth}, \& {Grant}}]{2023LRSP...20....1J}
{Jess}, D.~B., {Jafarzadeh}, S., {Keys}, P.~H., {et~al.} 2023, Living Reviews in Solar Physics, 20, 1

\bibitem[{Jess {et~al.}(2010)Jess, Mathioudakis, Christian, Keenan, Ryans, \& Crockett}]{jess_rosa_2010}
Jess, D.~B., Mathioudakis, M., Christian, D.~J., {et~al.} 2010, \solphys, 261, 363

\bibitem[{{Jess} {et~al.}(2015){Jess}, {Morton}, {Verth}, {Fedun}, {Grant}, \& {Giagkiozis}}]{2015SSRv..190..103J}
{Jess}, D.~B., {Morton}, R.~J., {Verth}, G., {et~al.} 2015, \ssr, 190, 103

\bibitem[{{Jess} {et~al.}(2020){Jess}, {Snow}, {Houston}, {Botha}, {Fleck}, {Krishna Prasad}, {Asensio Ramos}, {Morton}, {Keys}, {Jafarzadeh}, {Stangalini}, {Grant}, \& {Christian}}]{2020NatAs...4..220J}
{Jess}, D.~B., {Snow}, B., {Houston}, S.~J., {et~al.} 2020, Nature Astronomy, 4, 220

\bibitem[{Jess {et~al.}(2017)Jess, Van~Doorsselaere, Verth, Fedun, Krishna~Prasad, Erdélyi, Keys, Grant, Uitenbroek, \& Christian}]{jess_inside_2017}
Jess, D.~B., Van~Doorsselaere, T., Verth, G., {et~al.} 2017, \apj, 842, 59

\bibitem[{{Keys} {et~al.}(2019){Keys}, {Morton}, {Jess}, {Verth}, {Grant}, {Mathioudakis}, {MacKay}, {Doyle}, {Christian}, \& {Keenan}}]{keys_pores_2018}
{Keys}, P., {Morton}, R., {Jess}, D., {et~al.} 2019, ESS Open Archive eprints, 105, essoar.10500643

\bibitem[{Lagg {et~al.}(2010)Lagg, Solanki, Riethmüller, Martínez~Pillet, Schüssler, Hirzberger, Feller, Borrero, Schmidt, del Toro~Iniesta, Bonet, Barthol, Berkefeld, Domingo, Gandorfer, Knölker, \& Title}]{lagg_fully_2010}
Lagg, A., Solanki, S.~K., Riethmüller, T.~L., {et~al.} 2010, \apjl, 723, L164

\bibitem[{{Linnell Nemec} \& {Nemec}(1985)}]{1985AJ.....90.2317L}
{Linnell Nemec}, A.~F. \& {Nemec}, J.~M. 1985, \aj, 90, 2317

\bibitem[{{Morton} {et~al.}(2013){Morton}, {Verth}, {Fedun}, {Shelyag}, \& {Erd{\'e}lyi}}]{2013ApJ...768...17M}
{Morton}, R.~J., {Verth}, G., {Fedun}, V., {Shelyag}, S., \& {Erd{\'e}lyi}, R. 2013, \apj, 768, 17

\bibitem[{{Muller} {et~al.}(2007){Muller}, {Hanslmeier}, \& {Salda{\~n}a-Mu{\~n}oz}}]{2007A&A...475..717M}
{Muller}, R., {Hanslmeier}, A., \& {Salda{\~n}a-Mu{\~n}oz}, M. 2007, \aap, 475, 717

\bibitem[{{Nisenson} {et~al.}(2003){Nisenson}, {van Ballegooijen}, {de Wijn}, \& {S{\"u}tterlin}}]{2003ApJ...587..458N}
{Nisenson}, P., {van Ballegooijen}, A.~A., {de Wijn}, A.~G., \& {S{\"u}tterlin}, P. 2003, \apj, 587, 458

\bibitem[{Pesnell {et~al.}(2012)Pesnell, Thompson, \& Chamberlin}]{pesnell_solar_2012}
Pesnell, W.~D., Thompson, B.~J., \& Chamberlin, P.~C. 2012, \solphys, 275, 3

\bibitem[{Riethmüller {et~al.}(2014)Riethmüller, Solanki, Berdyugina, Schüssler, Martínez~Pillet, Feller, Gandorfer, \& Hirzberger}]{riethmuller_comparison_2014}
Riethmüller, T.~L., Solanki, S.~K., Berdyugina, S.~V., {et~al.} 2014, \aap, 568, A13

\bibitem[{Roberts(2019)}]{roberts_2019}
Roberts, B. 2019, MHD Waves in the Solar Atmosphere (Cambridge University Press)

\bibitem[{{Rutten}(2020)}]{rutten_solar_2020}
{Rutten}, R. 2020, in Solar Magnetic Variability and Climate, 29

\bibitem[{Schou {et~al.}(2012)Schou, Scherrer, Bush, Wachter, Couvidat, Rabello-Soares, Bogart, Hoeksema, Liu, Duvall, Akin, Allard, Miles, Rairden, Shine, Tarbell, Title, Wolfson, Elmore, Norton, \& Tomczyk}]{schou_design_2012}
Schou, J., Scherrer, P.~H., Bush, R.~I., {et~al.} 2012, \solphys, 275, 229

\bibitem[{Stangalini {et~al.}(2013)Stangalini, Berrilli, \& Consolini}]{stangalini_spectrum_2013}
Stangalini, M., Berrilli, F., \& Consolini, G. 2013, \aap, 559, A88

\bibitem[{Stangalini {et~al.}(2014)Stangalini, Consolini, Berrilli, De~Michelis, \& Tozzi}]{stangalini_observational_2014}
Stangalini, M., Consolini, G., Berrilli, F., De~Michelis, P., \& Tozzi, R. 2014, \aap, 569, A102

\bibitem[{Stangalini {et~al.}(2021)Stangalini, Jess, Verth, Fedun, Fleck, Jafarzadeh, Keys, Murabito, Calchetti, Aldhafeeri, Berrilli, Del~Moro, Jefferies, Terradas, \& Soler}]{stangalini_novel_2021}
Stangalini, M., Jess, D.~B., Verth, G., {et~al.} 2021, \aap, 649, A169

\bibitem[{Steiner {et~al.}(1998)Steiner, Grossmann-Doerth, Knölker, \& Schüssler}]{steiner_dynamical_1998}
Steiner, O., Grossmann-Doerth, U., Knölker, M., \& Schüssler, M. 1998, \apj, 495, 468

\bibitem[{Stenflo(1973)}]{stenflo_magnetic-field_1973}
Stenflo, J.~O. 1973, \solphys, 32, 41

\bibitem[{{Stenflo}(1989)}]{stenflo_small-scale_1989}
{Stenflo}, J.~O. 1989, \aapr, 1, 3

\bibitem[{{Tomczyk} \& {McIntosh}(2009)}]{2009ApJ...697.1384T}
{Tomczyk}, S. \& {McIntosh}, S.~W. 2009, \apj, 697, 1384

\bibitem[{{Torrence} \& {Compo}(1998)}]{1998BAMS...79...61T}
{Torrence}, C. \& {Compo}, G.~P. 1998, Bulletin of the American Meteorological Society, 79, 61

\bibitem[{{van Ballegooijen} {et~al.}(1998){van Ballegooijen}, {Nisenson}, {Noyes}, {L{\"o}fdahl}, {Stein}, {Nordlund}, \& {Krishnakumar}}]{1998ApJ...509..435V}
{van Ballegooijen}, A.~A., {Nisenson}, P., {Noyes}, R.~W., {et~al.} 1998, \apj, 509, 435

\bibitem[{Van~Doorsselaere {et~al.}(2014)Van~Doorsselaere, Gijsen, Andries, \& Verth}]{vanDoorsselaere2014}
Van~Doorsselaere, T., Gijsen, S.~E., Andries, J., \& Verth, G. 2014, The Astrophysical Journal, 795, 18

\bibitem[{Van~Doorsselaere {et~al.}(2020)Van~Doorsselaere, Srivastava, Antolin, Magyar, Farahani, Tian, Kolotkov, Ofman, Guo, Arregui, De~Moortel, \& Pascoe}]{van_doorsselaere_coronal_2020}
Van~Doorsselaere, T., Srivastava, A.~K., Antolin, P., {et~al.} 2020, Space Science Reviews, 216, 140

\end{thebibliography}

\begin{appendix}

    \section{Statistical characterisation of the dataset}\label{apa}

\begin{figure}[h!]
        \centering
        \includegraphics[scale=0.99]{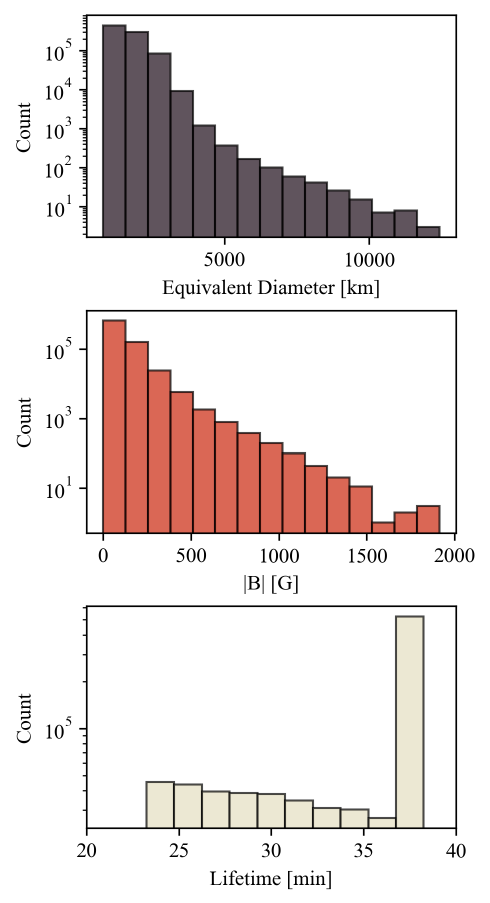}
        \caption{Physical properties of the considered magnetic structures. Top: Statistical distribution of the average equivalent diameter of the magnetic features detected in each temporal window. Middle: Statistical distribution of the average magnetic field of the elements in each temporal window, as inferred by the tracking. Bottom: Statistical distribution of the lifetime of the considered structures in this work.}
        \label{App1}
    \end{figure}

    In this section we provide the statistical characterisation of the elements considered in this work. As previously mentioned, we limited our analysis to 40-minute observational windows once every 3 days. We only considered elements with a lifetime of over 30 time steps ($\approx 22$ minutes). The first panel in Fig. \ref{App1} shows the statistical distribution of the average equivalent diameter of the elements considered in this work. The equivalent diameter was obtained by considering the magnetic structures to be perfectly round, thus exploiting the area in pixels as inferred by the tracking algorithm to compute the diameter. The distribution of |B| (G) is shown in the middle panel of Fig.\ref{App1}. It is worth noting that these distributions should only be considered representative of the magnetic concentrations considered in this work. Finally, the last panel shows the statistical distribution of the lifetime of the magnetic structures considered in this work.

    \section{Wavelet analysis}\label{apb}

    In this section we present the results of the wavelet analysis of the nearly 1 million small-scale magnetic elements included in our dataset. The procedure followed to carry out the global wavelet spectra of the time series is extensively described in \citet{1998BAMS...79...61T}. The pre-processing steps applied to each time series include subtracting the mean and normalising by standard deviation, as suggested in the aforementioned paper. This normalisation is essential to allow the comparison between the wavelet transforms at different scales.
    
\begin{figure}[h!]
        \centering
        \includegraphics{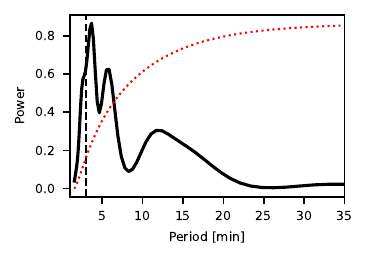}
        \caption{Randomly picked global wavelet spectrum. The dotted red curve represents the confidence level of 95\%. The dashed black line highlights the 3-minute period.}
        \label{App21}
    \end{figure}

    For each element in the dataset, we estimated the global wavelet spectrum of their horizontal velocity. In Fig. \ref{App21} we provide an example of the global wavelet spectrum of the considered magnetic structures in this work.

\begin{figure}[h!]
        \centering
        \includegraphics{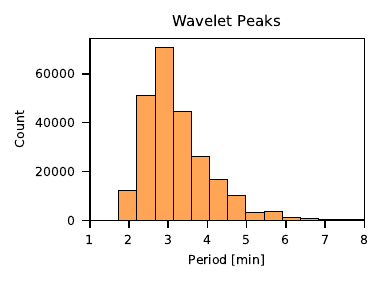}
        \caption{Statistical distribution of the highest peak over the 95\% significance level in each global wavelet spectrum.}
        \label{App22}
    \end{figure}

    In Fig. \ref{App22} we show the statistical distribution of the dominant period obtained by considering only the peaks with the highest power if it exceeded the 95\% confidence level. The confidence levels were estimated assuming red noise as the mean background spectrum. Additionally, the lag1 autocorrelation coefficient was computed for each time series. 

\end{appendix}

\end{document}